\documentclass{elsart4}

\usepackage{graphicx}
\usepackage{amssymb}
\usepackage{bm}

\begin{document}
\begin{frontmatter}

\title{Magnetic excitations in a new anisotropic Kagom\'e antiferromagnet}

\author[LLN]{J. Robert}
\author[LLN]{V. Simonet\corauthref{1}}
\corauth[1]{Corresponding Author: Email: simonet@grenoble.cnrs.fr}
\author[LLN]{B. Canals}
\author[LLN]{R. Ballou}
\author[LdC]{P. Bordet}
\author[LdC]{I. Gelard}
\author[LdC]{A. Ibanez}
\author[CRTBT]{P. Lejay}
\author[ILL]{J. Ollivier}
\author[ILL]{A. Stunault}
\address[LLN]{Laboratoire Louis N\'{e}el, CNRS, B.P. 166, 38 042
Grenoble Cedex 9, France}
\address[LdC]{Laboratoire de Cristallographie, CNRS, B.P. 166, 38
042 Grenoble Cedex 9, France}
\address[CRTBT]{Centre de Recherches des Tr\`es Basses
Temp\'eratures, CNRS, B.P. 166, 38 042 Grenoble Cedex 9, France}
\address[ILL]{Institut Laue-Langevin, BP 154, 38042 Grenoble Cedex,
France.}

\begin{abstract}
The Nd-langasite compound contains planes of magnetic Nd$^{3+}$
ions on a lattice topologically equivalent to a kagom\'e net. The
magnetic susceptibility does not reveal any signature of
long-range ordering down to 2 K but rather a correlated
paramagnetism with significant antiferromagnetic interactions
between the Nd and a single-ion anisotropy due to crystal field
effect. Inelastic neutron scattering on Nd-langasite powder and
single-crystal allowed to probe its very peculiar low temperature
dynamical magnetic correlations. They present unusual dispersive
features and are broadly localized in wave-vector $Q$ revealing a
structure factor associated to characteristics short
range-correlations between the magnetic atoms. From comparison
with theoretical calculations, these results are interpreted as a
possible experimental observation of a spin liquid state in an
anisotropic kagom\'e antiferromagnet.
\end{abstract}


\end{frontmatter}

The Heisenberg kagom\'e antiferromagnet is the archetypal example
of a highly frustrated magnetic 2-dimensional lattice, capable of
stabilizing a spin-liquid state. Extensive theoretical work was
devoted to the study of the peculiar nature of this spin liquid,
classically described by a non-magnetic highly degenerate
fluctuating ground state. Unfortunately, it is usually
destabilized by second-order perturbations, as well as by entropic
selection of soft modes via the "order by disorder" mechanism
\cite{KGMTh}. From the experimental side, very few examples of
ideal kagom\'e magnetic lattice were found in real systems which
were, moreover, often prone to non-stoichiometry. Among these, we
find the kagom\'e bilayers SCGO and BSZCGO \cite{bilayers}, the
jarosites \cite{jarosite}, and the natural volborthite
\cite{volb}, which stabilize non-conventional spin glasses, exotic
ordered phases, and show signatures of correlated paramagnetism
below the paramagnetic N\'eel temperature. All these are examples
of Heisenberg kagom\'e antiferromagnets. The case of anisotropic
kagom\'e antiferromagets, in which interesting new magnetic
behaviours are expected, has been much less studied theoretically
and was, up to now, still waiting for physical realizations.

The present study is devoted to a langasite compound, a family
better known for their application in the domain of
piezoelectricity \cite{Iwataki}. However, a thorough analysis of
their structure \cite{Mill} (space group P321) indicates that the
3e sites belonging to planes stacked perpendicular to the 3-fold
$c$-axis, form lattices with the same overall topology as the
kagom\'e one \cite{bordet}. In the studied Nd$_3$Ga$_5$SiO$_{14}$
compound, these sites are all occupied by the magnetic Nd$^{3+}$
ions, antiferromagnetically coupled to each other by
superexchange. Nd$^{3+}$, with electronic configuration 4f$^3$, is
expected to present strong anisotropy due to the crystal field
splitting of the fundamental multiplet $J$=9/2.

In the following, we report results of magnetization measurements
performed on a Quantum Design MPSMS SQUID magnetometer and of
inelastic neutron scattering measurements on powder sample and on
single-crystal. Large single-crystals of Nd-langasite were indeed
successfully grown by a floating zone method using an image
furnace, starting from a powder obtained through a solid state
reaction at 1420$^{\circ}$C in air \cite{lejay}. The powder
neutron scattering experiments were performed at the Institut
Laue-Langevin on the time-of-flight spectrometer IN5. The results
presented here were obtained with an incident wavelength of 4.5
\AA\ with a chopper speed of 12000 rpm and an energy resolution
(FWHM) of 100 $\mu$eV. Neutron scattering spectra were recorded at
2$~$K in the wave-vector $Q$ range [0.46, 2.48$~$\AA$^{-1}$] and
energy range [-197, +2.8 meV]. The dynamical magnetic correlations
and their localization in reciprocal space were more precisely
determined on a single-crystal at 2 K using the cold-neutron
three-axis spectrometer IN14 with fixed final energy of 4.66 meV
and energy resolution of 165 $\mu$eV.

\begin{figure}[t]
\includegraphics[scale=1]{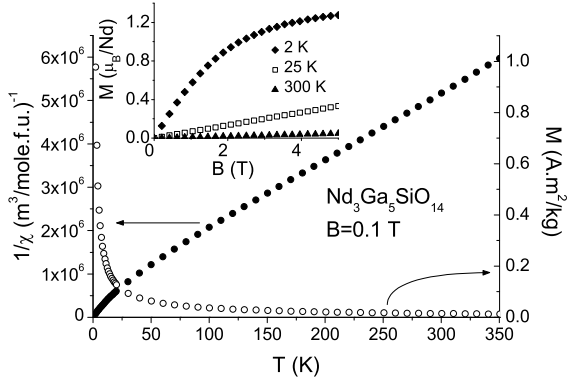}
\caption{Magnetization and inverse linear susceptibility measured
from 2 to 350 K in a 0.1 T magnetic field on Nd-langasite powder.
Inset : magnetization versus magnetic field at different
temperatures.} \label{Magn}
\end{figure}

As shown in Fig. \ref{Magn}, the thermal variation of the
magnetization measured on a powder sample under an applied
magnetic field of 0.1 T down to 2 K shows no anomaly nor any
thermomagnetic hysteresis that would indicate a transition towards
a long range order or a spin glass state. The inverse
susceptibility is linear down to 70 K before diving towards zero,
the linear part extrapolating to a negative intercept of the
temperature axis. The shape of the susceptibility is modulated by
the anisotropy. Its analysis, reported elsewhere \cite{bordet}, is
based on single-crystal measurements with magnetic field applied
parallel and perpendicular to the kagom\'e planes. The high
temperature analysis yields an effective moment $\mu_{\rm
eff}\approx$3.77 $\mu_{\rm B}$ close to the value of the Nd$^{3+}$
free ion and a paramagnetic N\'eel temperature $\theta$ of -52 K,
which confirms the existence of significant antiferromagnetic
interactions between the Nd$^{3+}$ ions. The fact that no long
range order is detected down to 2 K, a temperature well below the
$\theta$ value, shows that the compound is indeed frustrated and a
good candidate for a spin liquid phase. At high temperature, the
anisotropy of Nd-langasite is most probably described by coplanar
rotators lying in the kagom\'e planes. A change of the anisotropy
occurs at 33 K, the c axis becoming the magnetization one at lower
temperature, due to higher order anisotropy terms in the
crystalline electric field potential \cite{bordet}.

\begin{figure}
\includegraphics[scale=0.42]{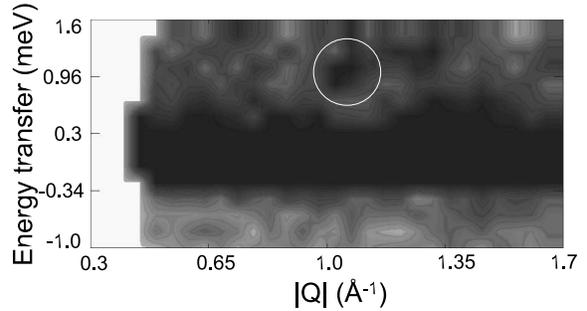}
\caption{Iso-intensity cut of the time-of-flight spectrum at 2 K
in Nd-langasite powder. Positive energy transfers are on the
neutron energy loss side. The darkest areas correspond to maximum
intensity : an horizontal stripe around the elastic position and a
localized one, spotted by the white circle.} \label{Tof}
\end{figure}

To characterize the magnetic excitations in the system, inelastic
neutron scattering measurements were carried out. The main
features revealed by the time-of-flight experiment are the low
levels of the $J$=9/2 multiplet splitted by the crystal field,
with a first intense one detected around 8.5 meV. In addition, a
much weaker signal could be detected, localized in modulus of $Q$
in contrast to the crystal field levels that are constant in $|Q|$
(neglecting the magnetic form factor). This signal is observed
between 0.8 and 1.2 meV and around 1.1 \AA$^{-1}$ (cf. Fig.
\ref{Tof}) \cite{robert}.

\begin{figure}[t]
\includegraphics[scale=0.75]{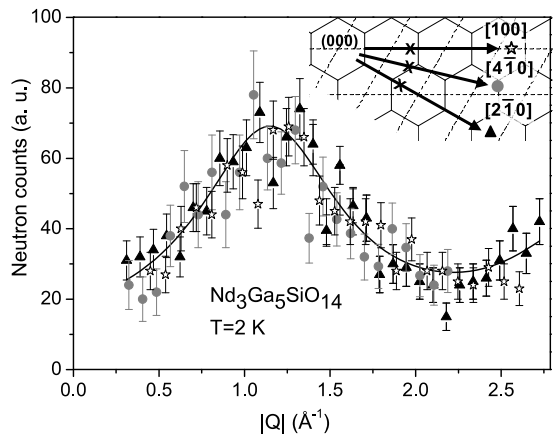}
\caption{Three-axis measurements on a Nd-langasite single-crystal:
$Q$-scans in the [100], [4$\bar 1$0] and [2$\bar 1$0] directions
at 0.85 meV drawn as a function of $|Q|$. The line is a lorentzian
based fit (see text). Right up corner : the spanning of the
reciprocal space, with $|a^*|$=$|b^*|$=0.899 \AA$^{-1}$, is
sketched and the neutron intensity maximum for the 3 directions is
indicated by a cross.} \label{3axe}
\end{figure}

A three-axis inelastic neutron scattering experiment on single
crystal was necessary in order to confidently measure and
characterize this small signal. The measurements were done in the
horizontal scattering plane containing the [100] and [010] axes.
Energy scans, performed at different points of the reciprocal
space, confirmed the presence of a small signal around 1 meV and
$Q$=1.1 \AA$^{-1}$. This signal mimics those of the calculated
static magnetic structure factors which suggests its magnetic
origin \cite{robert}. $Q$-scans were then performed at several
energies around 1 meV in different directions of the reciprocal
space. In Fig. \ref{3axe}, the results at an energy of 0.85 meV,
spanning the 15$^{\circ}$ rotated [100], [4$\bar 1$0] and [2$\bar
1$0] directions, are reported. The spectra for the 3 directions
are very similar yielding a ring-shaped maximum of intensity at
around 1.15 \AA$^{-1}$. Then, there is a minimum at 2.2 \AA$^{-1}$
and a second weak maximum rising at larger $Q$ values (at least
for the [2$\bar 1$0] direction). This underlines that the magnetic
intensity pattern is not equally distributed in all the BZ.

Quantitatively, the main peak was fitted with a lorentzian,
multiplied by the square of the Nd$^{3+}$ magnetic form factor.
This analysis yielded a HWHM of 0.49 \AA$^{-1}$, i.e. a very short
correlation length of 2 \AA, smaller than the distance between two
Nd (4.2 \AA). The full analysis of the magnetic $Q$ distribution
as a function of energy shows a complex behavior, which is
detailed elsewhere \cite{robert}. A second peak, whose position
varies, is found at certain energies. The first peak position also
changes slightly, which indicates a significant dispersion of the
magnetic scattering with $Q$ in the range 0.5 to 1 meV. The life
time of these magnetic fluctuations, estimated from the half width
of the energy response at constant $Q$, is of the order of 1.4
10$^{-12}~$s.

An interpretation of these results can be obtained in the light of
both experimental and theoretical works. Calculations were
performed using several models of spin-spin correlation functions
on the antiferromagnetic kagom\'e lattice obtained with the real
crystallographic positions of the Nd$^{3+}$ ions \cite{robert}.
The calculations give the static magnetic structure factor for the
spin liquid kagom\'e phase, which is characteristic of the low
energy spin fluctuations. As expected, the $Q$-distribution of
magnetic intensity is therefore in best agreement with the
experimental results at the lowest investigated energy of 0.5 meV
\cite{robert}. Actually, when first neighbors only are considered
whatever the spin dimension, an empty first BZ and a ring of
maximum intensity around 0.9 \AA$^{-1}$ are found. This magnetic
structure factor, based on a disordered state, is for instance
compatible with short-range magnetic correlations between 3
Nd$^{3+}$ first neighbors on a triangle forming a non magnetic
state in the XY and Heisenberg cases. These non magnetic states
were already invoked to interpret the $Q$ pattern of the dynamical
correlations obtained in previous inelastic neutron scattering
experiments on single-crystal compounds containing frustrated
pyrochlore lattice of magnetic atoms : the itinerant
Y$_{0.97}$Sc$_{0.03}$Mn$_2$ \cite{ballou} and the insulating
ZnCr$_2$O$_4$ \cite{lee}. The $Q$ patterns, measured in these
compounds and interpreted as originating from spin liquid
correlations, are indeed close to the present one. The similarity
can be related to the fact that a cut perpendicular to the cube
diagonal of the pyrochlore lattice leads to the kagom\'e one. In
addition, the very short correlation lengths found in Nd-langasite
and in Y$_{0.97}$Sc$_{0.03}$Mn$_2$ \cite{ballou} are thought to
come from the high degeneracy of the magnetic modes which is
another signature of a spin -liquid state in a frustrated
antiferromagnet. However, for the Y$_{0.97}$Sc$_{0.03}$Mn$_2$
compound, the life time of the spin correlations was found much
shorter than in the present case and no dispersion was observed.
The understanding of these differences and the thorough
characterization of the $Q$-distribution of the magnetic intensity
must now be tackled with realistic calculations taking into
account the anisotropy of the system.

In conclusion, the analysis of the static magnetic properties and
of the dynamical magnetic correlations of Nd-langasite suggests
that this compound could be the first example of a spin liquid
state stabilized in an anisotropic kagom\'e antiferromagnet.


\begin{thebibliography}{00}
\bibitem{KGMTh} D. A. Huse {\it et al.}, Phys. Rev. B {\bf 45},
7536 (1992); A.B.Harris {\it et al.}, Phys. Rev. B {\bf 45}, 2899
(1992); J. N. Reimers {\it et al.}, Phys. Rev. B {\bf 48}, 9539
(1993); P. W. Leung {\it et al.}, Phys. Rev. B {\bf 47}, 5459
(1993); Ch. Waldtmann {\it et al.}, Eur. Phys. J. B {\bf 2}, 501
(1998);D. A. Garanin, and B. Canals, Phys. Rev. B {\bf 59}, 443
(1999).
\bibitem{bilayers} L. Limot {\it et al.}, Phys. Rev. B {\bf 65}, 144447
(2002); D. Bono {\it et al.}, Phys. Rev. Lett. {\bf 92}, 217202
(2004).
\bibitem{jarosite} A. S. Wills, Can. J. Phys. {\bf 79}, 1503
(2001); D. Grohol {\it et al.}, Nature Materials {\bf 4}, 323
(2005).
\bibitem{volb} A. Fukaya {\it et al.}, Phys. Rev. Lett.
{\bf 91}, 207603 (2003); F. Bert {\it et al.}, cond-mat/0507250.
\bibitem{Iwataki} T. Iwataki, H. Ohsato, K. Tanaka, H. Morokoshi, J. Sato, and K. Kawasaki, J. Eur. Cer. Soc. {\bf 21}
1409 (2001).
\bibitem{Mill} B. V. Mill, A. V. Butashin, G. G. Kodzhabagyan, E. L. Belokonova, and
N. V. Belov, Dokl. Akad. Nauk SSSR {\bf 264} (6) 1395 (1982).
\bibitem{bordet} P. Bordet {\it et al.}, submitted to J. Phys.: Condens. Matter (2005).
\bibitem{lejay} P. Lejay {\it et al.}, in preparation.
\bibitem{robert} J. Robert {\it et al.}, in preparation.
\bibitem{ballou} R. Ballou {\it et al.}, Phys. Rev. Lett. {\bf 76}, 2125
(1996); R. Ballou, Can. J. Phys. {\bf 79}, 1475 (2001).
\bibitem{lee} S. -H. Lee {\it et al.}, Nature {\bf 418}, 856
(2002).

\end{thebibliography}
\end{document}